\documentclass[aps,prl,twocolumn,amsfonts,amssymb,amsmath,floatfix,10pt,superscriptaddress]{revtex4-1}
\usepackage[english]{babel}
\usepackage{graphicx}
\usepackage{ulem}
\usepackage[hidelinks]{hyperref}
\usepackage{siunitx}

\DeclareSIUnit{\Erecoil}{E_r}
\DeclareSIUnit{\Tfermi}{T_F}
\DeclareSIUnit{\kbrillouin}{k_{BZ}}
\DeclareSIUnit{\ab}{a_{0}}
\DeclareSIUnit{\abb}{a_{bb}}
\DeclareSIUnit{\abf}{a_{bf}}

\begin{document}

\title{Quantum superhet based on microwave-dressed Rydberg atoms}
\author{Mingyong Jing} 
\thanks{These authors contributed equally to this work.}
\affiliation{State Key Laboratory of Quantum Optics and Quantum Optics Devices, Institute of Laser Spectroscopy, Shanxi University, Taiyuan, Shanxi 030006, China}
\affiliation{Collaborative Innovation Center of Extreme Optics, Shanxi University, Taiyuan, Shanxi 030006, China}

\author{Ying Hu} 
\thanks{These authors contributed equally to this work.}
\affiliation{State Key Laboratory of Quantum Optics and Quantum Optics Devices, Institute of Laser Spectroscopy, Shanxi University, Taiyuan, Shanxi 030006, China}
\affiliation{Collaborative Innovation Center of Extreme Optics, Shanxi University, Taiyuan, Shanxi 030006, China}

\author{Jie Ma}
\affiliation{State Key Laboratory of Quantum Optics and Quantum Optics Devices, Institute of Laser Spectroscopy, Shanxi University, Taiyuan, Shanxi 030006, China}
\affiliation{Collaborative Innovation Center of Extreme Optics, Shanxi University, Taiyuan, Shanxi 030006, China}

\author{Hao Zhang}
\affiliation{State Key Laboratory of Quantum Optics and Quantum Optics Devices, Institute of Laser Spectroscopy, Shanxi University, Taiyuan, Shanxi 030006, China}
\affiliation{Collaborative Innovation Center of Extreme Optics, Shanxi University, Taiyuan, Shanxi 030006, China}

\author{Linjie Zhang}
\thanks{zlj@sxu.edu.cn}
\affiliation{State Key Laboratory of Quantum Optics and Quantum Optics Devices, Institute of Laser Spectroscopy, Shanxi University, Taiyuan, Shanxi 030006, China}
\affiliation{Collaborative Innovation Center of Extreme Optics, Shanxi University, Taiyuan, Shanxi 030006, China}

\author{Liantuan Xiao}
\thanks{xlt@sxu.edu.cn}
\affiliation{State Key Laboratory of Quantum Optics and Quantum Optics Devices, Institute of Laser Spectroscopy, Shanxi University, Taiyuan, Shanxi 030006, China}
\affiliation{Collaborative Innovation Center of Extreme Optics, Shanxi University, Taiyuan, Shanxi 030006, China}

\author{Suotang Jia}
\affiliation{State Key Laboratory of Quantum Optics and Quantum Optics Devices, Institute of Laser Spectroscopy, Shanxi University, Taiyuan, Shanxi 030006, China}
\affiliation{Collaborative Innovation Center of Extreme Optics, Shanxi University, Taiyuan, Shanxi 030006, China}

\maketitle

\textbf{The highly sensitive, phase- and frequency-resolved detection of microwave electric fields is of central importance for diverse fields ranging from astronomy~\cite{Spitler2016,Shannon2018}, remote sensing~\cite{Alsdorf2000,Rignot2008}, communication~\cite{Rohde2017} and microwave quantum technology~\cite{Maeda2005,Bienfait2017}. However, present quantum sensing of microwave electric fields primarily relies on atom-based electrometers~\cite{Sedlacek2012,Fan2015} only enabling amplitude measurement. Moreover, the best sensitivity of atom-based electrometers is limited by photon shot noise to few $\mu$Vcm$^{-1}$Hz$^{-1/2}$~\cite{Kumar2017,Kumar2017a}: While going beyond is in principle possible by using squeezed light or Schr\"odinger-cat state, the former is very challenging for atomic experiments while the latter is feasible in all but very small atomic systems~\cite{Facon2016}. Here we report a novel microwave electric field quantum sensor termed as quantum superhet, which, for the first time, enables experimental measurement of phase and frequency, and makes a sensitivity few tens of nVcm$^{-1}$Hz$^{-1/2}$ readily accessible for current experiments. This sensor is based on microwave-dressed Rydberg atoms and tailored optical spectrum, with very favorable scalings on sensitivity gains. We can experimentally achieve a sensitivity of $55$ nVcm$^{-1}$Hz$^{-1/2}$, with the minimum detectable field being three orders of magnitude smaller than existing quantum electrometers. We also measure phase and frequency, being able to reach a frequency accuracy of few tens of $\mu$Hz for microwave field of just few tens of nVcm$^{-1}$. Our technique can be also applied to sense electric fields at terahertz or radio frequency. This work is a first step towards realizing the long sought-after electromagnetic-wave quantum sensors with quantum projection noise limited sensitivity, promising broad applications such as in radio telescope, terahertz communication~\cite{Koenig2013,Nagatsuma2016} and quantum control.}

Quantum sensing harnesses highly coherent and well controlled quantum systems to measure weak signals with unprecedented sensitivity and precision~\cite{Degen2017}. In particular, Rydberg atom provides a platform allowing high-sensitivity quantum sensing of microwave (MW) electric field~\cite{Fan2015,Degen2017}. This has been highlighted by recent experiments, with the development of self-calibrated Rydberg-atom quantum electrometers outperforming the classical counterpart~\cite{Sedlacek2012,Sedlacek2013,Kumar2017,Kumar2017a,Holloway2014,Holloway2017,Cox2018}, and sensing techniques with sub-wavelength spatial resolution enabled by the optical readout~\cite{Boehi2012,Wade2017}. 
\begin{figure}[tb]
\centering
\includegraphics[width= 1\columnwidth]{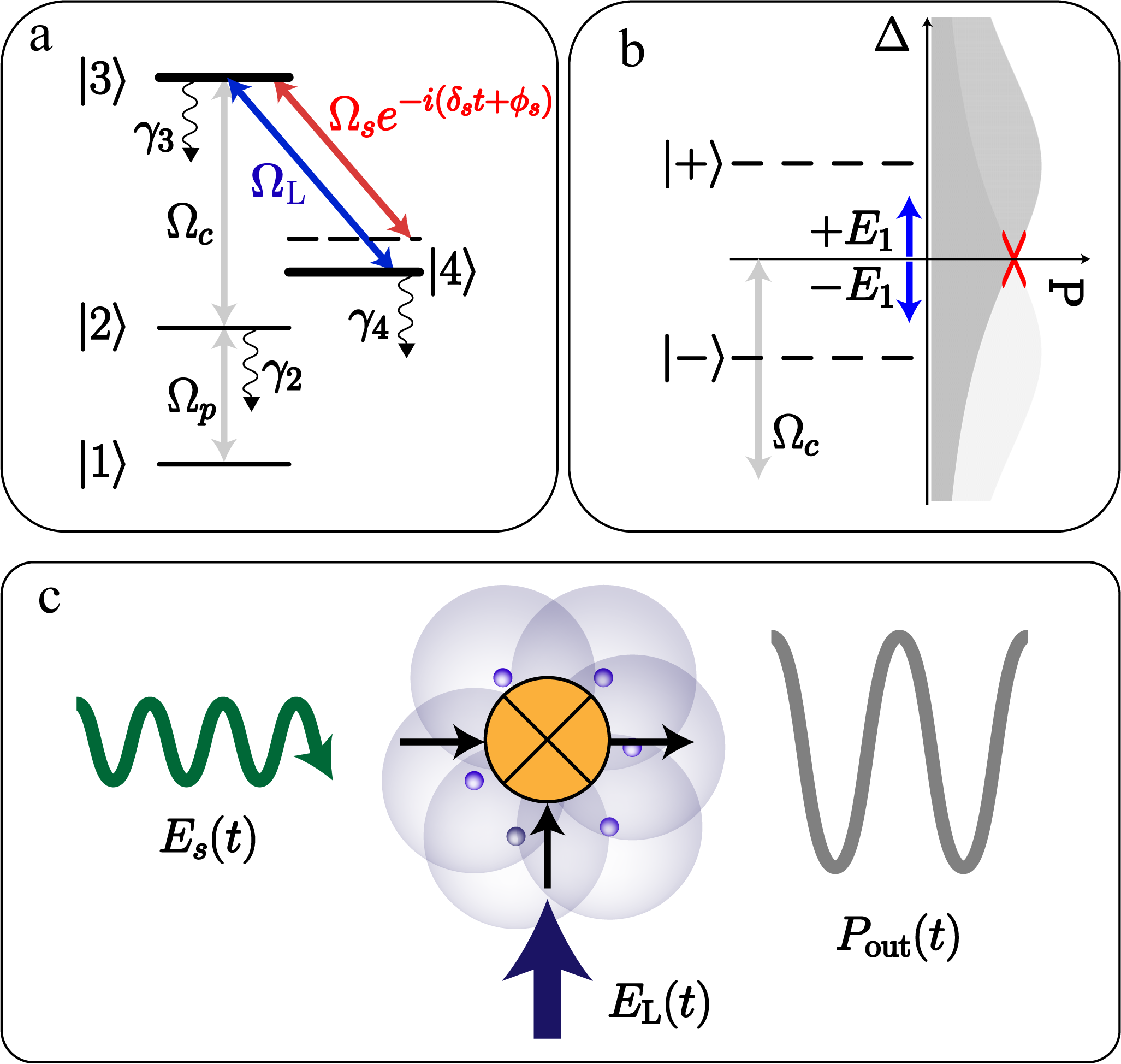}
\caption{Atom-based quantum superhet. (a) Setup:  states $|1\rangle$, $|2\rangle$ and Rydberg state $|3\rangle$ are resonantly coupled by a probe ($\Omega_p$) and control fields ($\Omega_c$), respectively. A local MW electric field (blue) is resonant with Rydberg transition $|3\rangle-|4\rangle$. A weak signal MW electric field (red) yields a coupling $\Omega_se^{-i(\delta_st+\phi_s)}$, with a phase $\phi_s$ and frequency detuning $\delta_s$ relative to the local field. State $|i\rangle$ ($i=2,3,4$) has a decay rate $\gamma_i$.  (b) MW-dressed Rydberg states and linear EIT spectrum near zero laser detuning $\Delta=0$:  two dressed states $|\pm\rangle$ are created for $\Omega_{\textrm{L}}\neq 0$ and $\Omega_s=0$, inducing AT-splitting of EIT lines (gray). For $\Omega_s\neq 0$, $|\pm\rangle$ acquires an energy correction $\pm E_1=\Omega_s\cos(\delta_st+\phi_s)/2$ shifting EIT lines outward. When $\Omega_{\textrm{L}}\sim \Gamma_{\textrm{EIT}}$, both EIT lines are linear near $\Delta=0$. (c) Operating principles: a MW electric field $E_s(t)$ is detected as optical signal $P_{\textrm{out}}(t)$, with downshifted frequency, original phase and linearly-enhanced amplitude, reminiscent of the superhet based on nonlinear mixing familiar in the context of electromagnetic-wave detections. } \label{fig:1}
\end{figure} 

 \begin{figure*}[hbt!]
\centering
\includegraphics[width=0.96\textwidth]{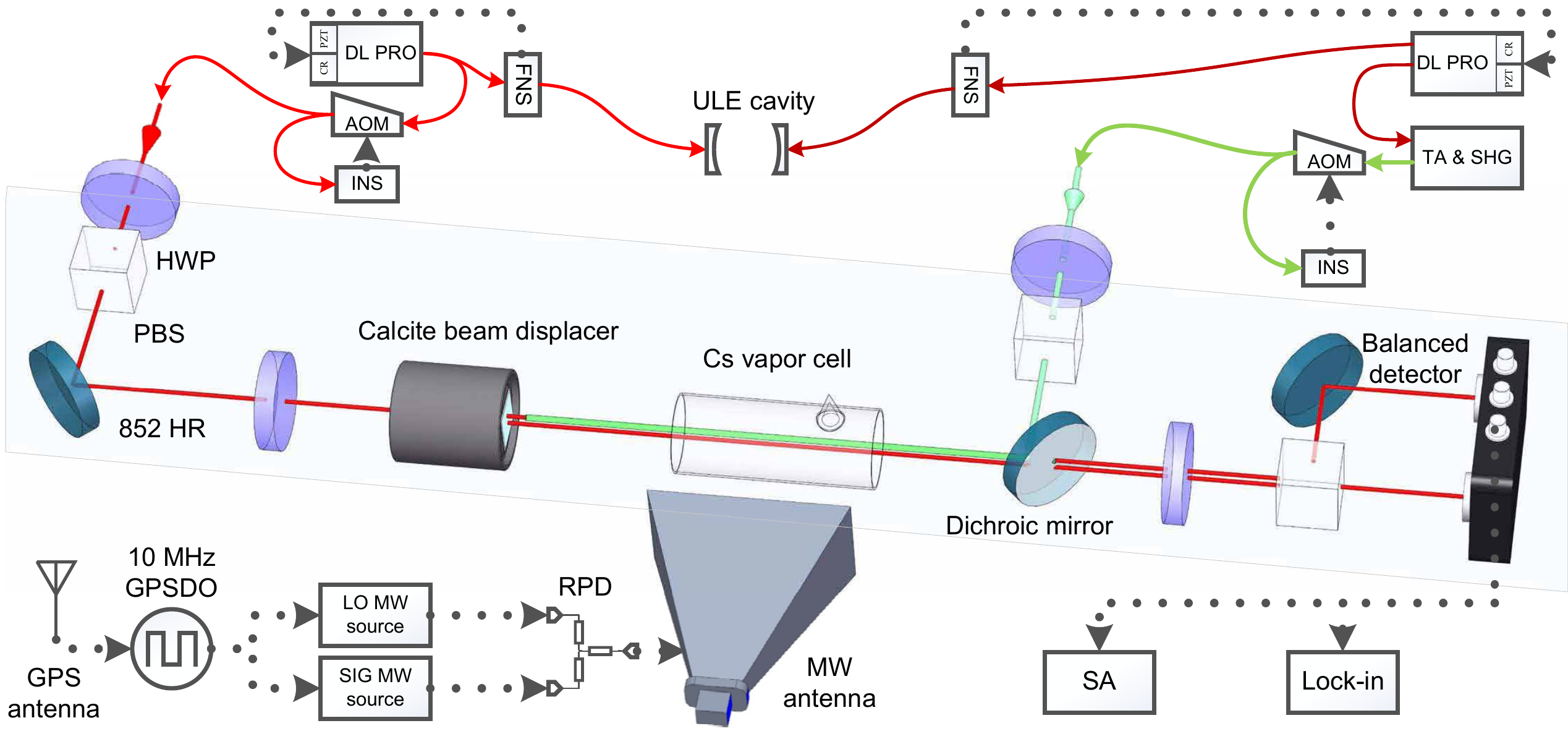}
\caption{Overview of the experimental setup including quantum superhet and detection systems (Supplementary material). We have also used the following notations: (1) HWP: half wave plate, (2) PBS: polarizing beam splitter, (3) HR: dielectric mirror,  (4) GPSDO: GPS disciplined oscillator with Rubidium timebase, (5) RPD: 2-way microwave resistive power divider, (6) SA: FFT spectrum analyzer, (7) Lock-in: lock-in amplifier, (8) ULE: ultra-low expansion (ULE) glass cavity, (9) AOM: The double-pass acousto-optic modulators which shift the frequencies of the probe and coupling lights to atomic resonances, (10) INS: intensity noise server.}\label{fig:2}
\end{figure*}  

\begin{figure*}[tb!]
\centering
\includegraphics[width=0.96\textwidth]{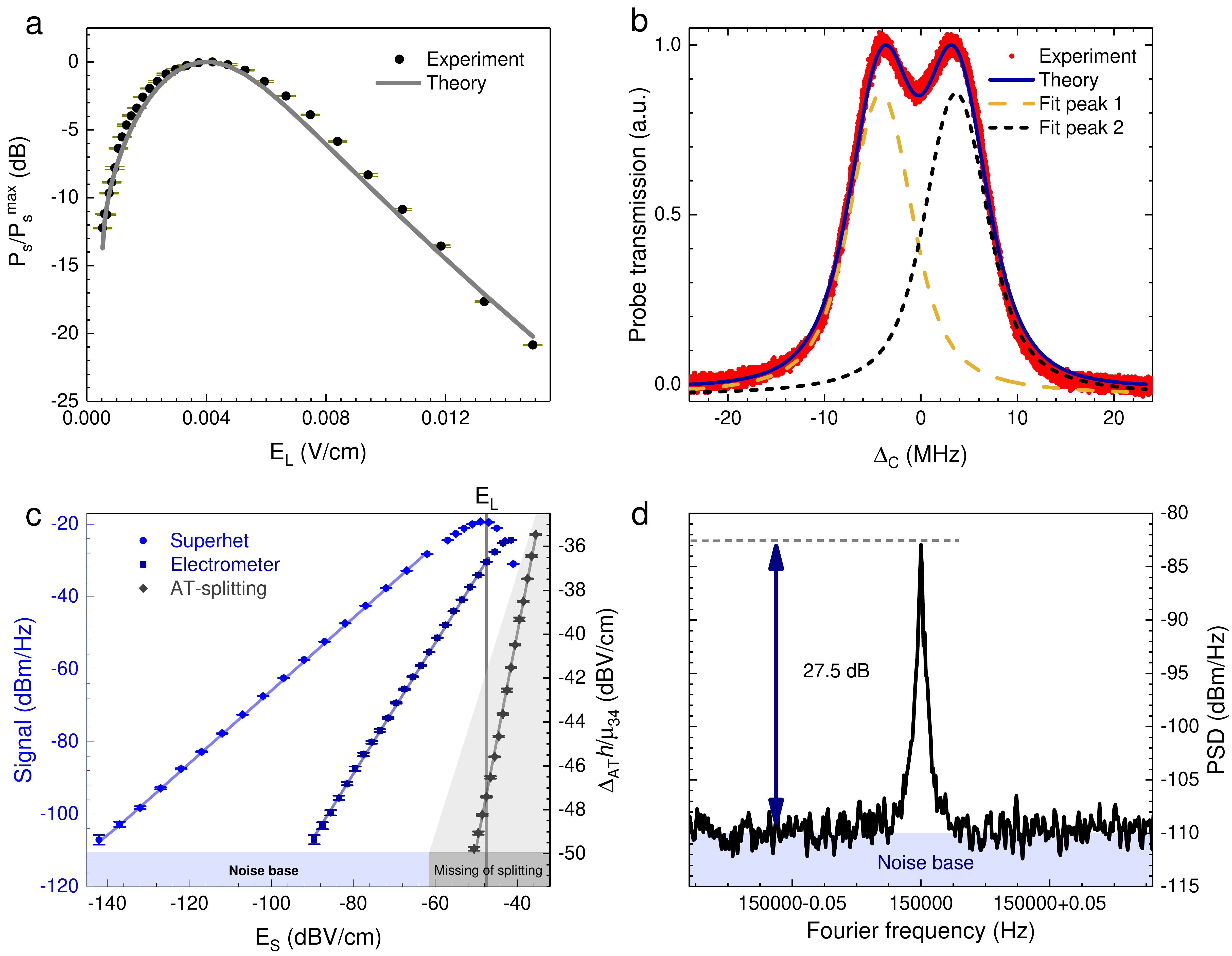}
\caption{\label{fig:3} Experimental measurement of MW electric field amplitude. (a) $P_s/P_s^{\textrm{max}}$ as a function of local field amplitude $E_{\textrm{L}}$ [c.f. Eq.~(\ref{Eq:A})], where $P_s^{\textrm{max}}$ denotes the maximum EIT transmission. Experimental (black dotted) and theoretical results (gray solid) are shown. (b) EIT signal as a function of control laser detuning $\Delta_c$, when $\Omega_{\textrm{L}}$ corresponds to the peak in (a). Experimental (red dotted) and theoretical results (blue solid) are shown. Dashed curves present the multi-peak fit of the experiment data. (c) Slope detection of quantum superhet vs. nonlinear detection of atomic electrometer: $P_\textrm{out}(E_s)$ [c.f. Eq.~(\ref{eq:Pout})] for quantum superhet (light blue with filled circles), and $\Delta P(E_s)=P(E_s)-P(0)$ for atomic electrometer~\cite{Sedlacek2012}(dark blue with filled square). Solid curves are linear fittings. The gray dotted curve depicts AT splitting as a function of $E_s$ with corresponding fitting (grad solid). (d) Fourier analysis of $P_{\textrm{out}}(t)$ for $E_\textrm{L}=55$ nVcm$^{-1}$. In (a), (b) and (c), spectrum analzyer has RBW of $1$ Hz. In (b), (c) and (d), we have fixed $E_\textrm{L}=3.0$ mVcm$^{-1}$ corresponding to $\Omega_{\textrm{L}}=7.9$ MHz. In all plots, the error bar is obtained from the statistics of data from 10 experiments.}\label{fig:3}
\end{figure*}  

Two central tasks in quantum sensing of time-varying electric fields are (1) to diagnose the phase and frequency of weak fields, which is at the basis of, as a paradigmatic example, radar detections~\cite{Doviak2012}; and (2) to detect small field amplitude with ultrahigh sensitivity. However, existing atom-based MW electric field quantum sensors only allow to probe field amplitude from the optical readout~\cite{Sedlacek2012,Sedlacek2013,Kumar2017,Kumar2017a,Holloway2014,Holloway2017,Cox2018,Boehi2012,Wade2017}, severely limiting their practical applications in actual MW detections. In the quest to measure phase and frequency, while there are proposals involving complicated techniques~\cite{Fan2015,Pandey2018}, demonstrations with realistic tools in atom-based experiments have remained elusive. Further, the sensitivity of current atom-based electrometers have approached the photon shot noise limit, with $3$ $\mu$Vcm$^{-1}$Hz$^{-1/2}$ at best even when assisted with sophisticated approaches~\cite{Kumar2017}. To beat this limit is very demanding as it requires non-classical squeezed light in relevant atomic setups. The origin for such difficulty to improve the sensitivity of existing atom-based MW electrometers is that they only realize \textit{nonlinear} detection when the MW signal is so weak that corresponding Rabi frequency is smaller than the linewidth of optical spectrum, leading to (approximately) unfavorable $\propto \sigma^{1/2}$ scaling of sensitivity with classical-noise induced error $\sigma$. Alternative route to high sensitivity exploits non-classical Schr\"{o}dinger-cat states, as recently demonstrated~\cite{Facon2016}, but relevant experiments have been limited to systems with a small number of atoms due to the difficulty in preparing cat-states. Developing new quantum sensing schemes to overcome both above limitations is currently an outstanding challenge. 

Here we use a strong local MW to dress Rydberg states and tailor the electromagnetically induced transparency~\cite{Fleischhauer2005} (EIT) spectrum, thus realizes a novel quantum sensor reminiscent of the superhet in conventional electromagnetic-wave detection architectures~\cite{Rohde2017,Doviak2012}; see Fig.~\ref{fig:1}. As we will show, quantum superhet not only allows detecting phase and frequency of microwave electric fields, it also realizes very favorable $\propto \sigma$ scaling of sensitivity, hence makes a remarkably high sensitivity $\lesssim 55$ nVcm$^{-1}$Hz$^{-1/2}$ readily accessible to experiments even without non-classical resources, which further benefit measurement accuracies of frequency and phase. 

We illustrate the operating principles of quantum superhet based on the setup in Fig.~\ref{fig:1} (a). The key new ingredient is a strong local MW electric field resonant with Rydberg transitions, with Rabi frequency $\Omega_\textrm{L}$ corresponding to a field amplitude $E_{\textrm{L}}$. Both the probe and coupling beams are resonant with corresponding atomic transitions. In the rotating wave approximation, a signal MW field with amplitude $E_{\textrm{s}}$ leads to a coupling $\Omega_s e^{-i(\delta_st+\phi_s)}$ between Rydberg states, where the frequency detuning $\delta_s$ and phase $\phi_s$ are measured relative to the local field. We will be interested in the regime $\delta_s\ll \Gamma_{\textrm{EIT}}$ with $\Gamma_{\textrm{EIT}}$ representing typical EIT linewidth, thus quantum superhet is in the instantaneous steady state within the adiabatic approximation. 

In quantum superhet, a MW signal has its entire information encoded in the first order energy shift of MW-dressed Rydberg states [Fig.~\ref{fig:1} (b)]. When $\Omega_s=0$, the strong on-resonant local MW field results in two dressed states $|\pm\rangle$ energetically separated by $\Omega_{\textrm{L}}$, which are respectively the symmetric and antisymmetric superpositions of two bare Rydberg states $|3\rangle$ and $|4\rangle$. When perturbed by $\Omega_s e^{-i(\delta_st+\phi_s)}$ with $\Omega_s/\Omega_\textrm{L}\ll 1$, $|\pm\rangle$ acquires an instantaneous first-order energy shift $\pm E_1=\pm \Omega_s\cos(\delta_st+\phi_s)/2$, which preserves original phase and frequency detuning.

In detecting $E_1$, we rely on using $\Omega_{\textrm{L}}$ as a control knob to realize a \textit{linear} EIT spectrum of dressed Rydberg atoms in close vicinity of zero laser detuning [Fig.~\ref{fig:1} (b)]. When $\Omega_s=0$, the EIT spectrum of MW-dressed Rydberg atoms shows familiar Autler-Townes (AT) splitting~\cite{Fleischhauer2005} of the EIT peak, with the separation of one EIT line from the other depending on $\Omega_{\textrm{L}}$. We tune $\Omega_{\textrm{L}}$ in such a way that both EIT lines become \textit{linear} (red lines) near zero laser detuning, which can occur for ${\Omega_{\textrm{L}}\sim \Gamma_{\textrm{EIT}}}$. Consequently, when an energy correction $\pm E_1$ shifts both EIT lines outwards, this shift transforms linearly into changes of on-resonance EIT signals with the \textit{maximum} amplification rate, thus realizes the highly desired slope detection~\cite{Degen2017} for $\Omega_s\ll \Gamma_{\textrm{EIT}}$.

Thus a MW signal is directly measured as an optical readout represented by [Fig.~\ref{fig:1} (c)]:
\begin{equation}
P_\textrm{out}(t)=P(t)-\bar{P},\label{eq:Pout}
\end{equation}
where $P(t)$ and $\bar{P}$ denote EIT signals of MW-dressed atoms measured at zero laser detuning with and without $\Omega_s$. We obtain (Supplementary material):
\begin{equation}
P_\textrm{out}(t)=P_s\cos(\delta_s t+\phi_s), \label{eq:EIT}
\end{equation} 
where the amplitude is
\begin{equation}
P_s=\left(\frac{\alpha \bar{P}}{\Gamma}\right)\Omega_s=\left(\frac{\sqrt{2}\mu_r\alpha \bar{P}}{\hbar\Gamma}\right)E_s. \label{Eq:A}
\end{equation}
Here $\alpha\le 1$ denotes the ratio of photons participating the EIT process, $\Gamma=\tau_c^{-1}$ defines the coherence time $\tau_c$ of the quantum superhet intimately related to $\Gamma_{\textrm{EIT}}$, $\mu_r$ is the dipole moment associated with Rydberg transition. Above equations are ensured by $\Omega_s/\Omega_{\textrm{L}}\ll 1$ and $\alpha\Omega_s/\Gamma\ll 1$, and is generally valid for both cold and thermal atoms in experimentally realistic conditions. 

As we will show, quantum superhet has two significant advantages in MW electric field sensing: (1) For $\Omega_s\ll \Gamma_{\textrm{EIT}}$, the quantum superhet realizes slope detection with the benefit of favorable scalings, in particular $\propto \sigma$ of sensitivity. This dramatically improves the efficiency in the effort to improve sensitivity by reducing classical noise, contrasting to Rydberg-atom MW electrometers which instead detect $\Omega_s$ nonlinearly. (2) The frequency resolution of $\delta_s$ measurement does not depend on the coherence time $\tau_c$ but rather limited only by the stability of an external synchronization clock. 

\begin{figure}[tb]
\centering
\includegraphics[width= 1\columnwidth]{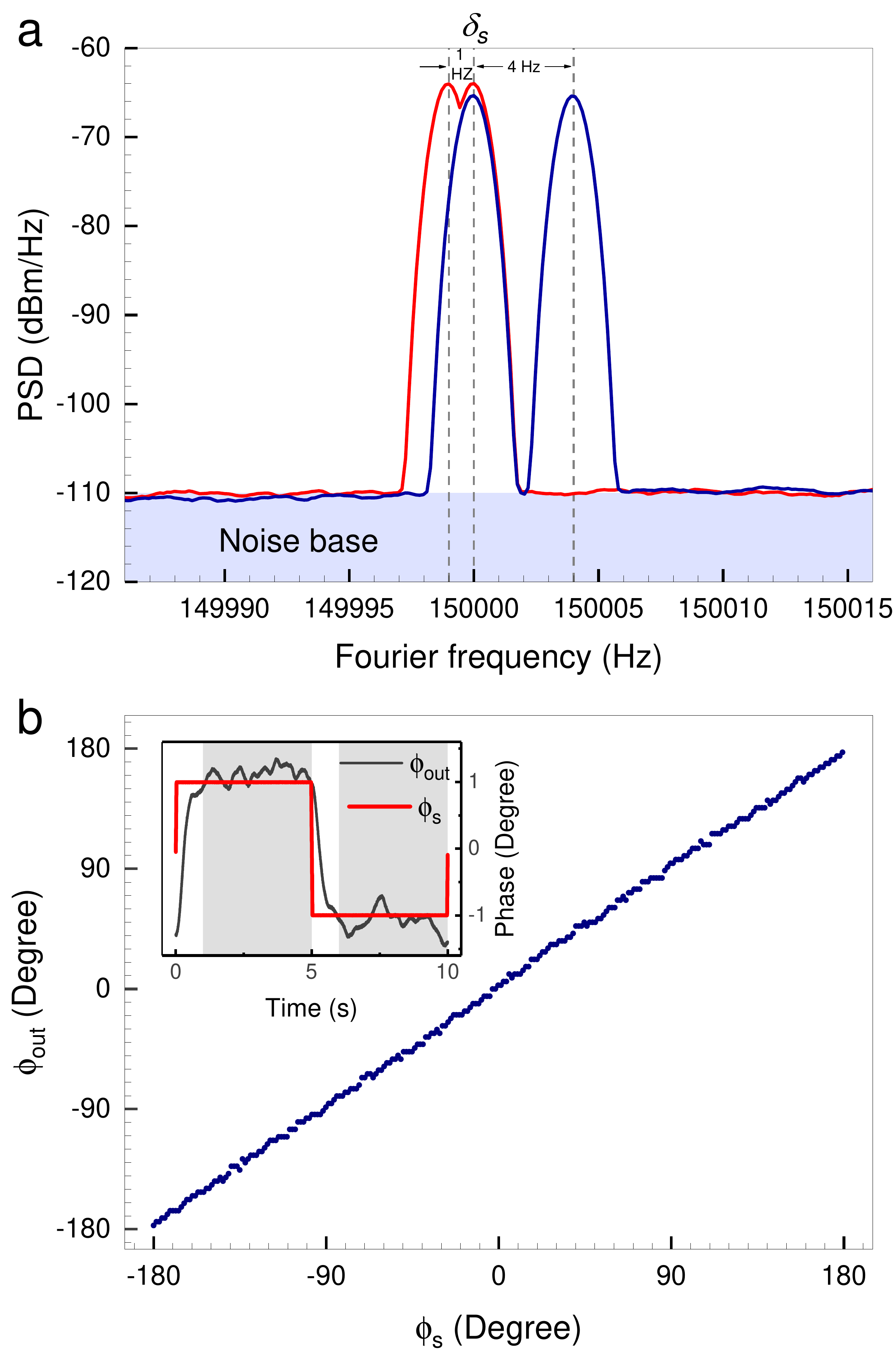}
\caption{Experimental measurement of frequency shift and phase of signal MW electric fields. (a) Experimental data for the Fourier transform of $P_{\textrm{out}}(t)$ in the frequency domain, so as to measure $\delta_d^i$ in a MW signal $\Omega_{\textrm{in}}(t)=\sum_{i=1,2} r_i\Omega_t e^{-i\delta^i_d t}e^{-i\delta_s t}$ (see main text). We have chosen $r_i\Omega_t=20$ kHz, $\delta_d^1=0$, while $\delta_d^2=1$ Hz (red) and $\delta_d^2=4$ Hz (blue). (b) Comparisons between the measured phase $\phi_{\textrm{out}}$ and the original phase $\phi_s$ in a signal $\Omega_s e^{-i(\delta_s t+\phi_s)}$. $\phi_{\textrm{out}}$ is obtained from the phase sensitive measurement of $P_{\textrm{out}}(t)$, $\Omega_s=20$ kHz. Inset: $\phi_{\textrm{out}}$ for a jump phase $\phi_s$ whose values varies by 1$^{\circ}$ stepwisely. Standard deviation of $\phi_{\textrm{out}}$ is obtained for datas sampled in the time interval in gray block (Supplementary material). In both (a) and (b), $\delta_s=150.000$ kHz. }\label{fig:4}
\end{figure}  

We experimentally implement quantum superhet using Cs atoms in a room-temperature vapor cell (Fig.~\ref{fig:2} and Supplementary material). Before exploring it to detect a signal, we optimize its sensitivity harnessing the controllability of $E_{\textrm{L}}$. Using a test signal with $\delta_s= 150.000$ kHz, we measure $P(t)$ for various $E_{\textrm{L}}$ by a spectrum analyzer with resolution bandwidth (RBW) of $1$ Hz. We obtain $\bar{P}$ from the time average of $P(t)$, which for $\Omega_s\ll \Omega_{\textrm{L}}$ provides the EIT signal of MW-dressed atoms ($\Omega_s=0$). Figure~\ref{fig:3} (a) shows that the best sensitivity occurs when $E_{\textrm{L}}=3.0$ mVcm$^{-1}$, corresponding to $\Omega_{\textrm{L}}=7.9$ MHz. A theoretical estimation (Supplementary material) is also shown (solid gray trace), which agrees well with the experiment. Figure~\ref{fig:3} (b) presents the AT-splitting measurement for $\Omega_{\textrm{L}}=7.9$ MHz ($E_s=0$) where the coupling laser is scanned. The experimental data (red trace, which agrees with theoretical results shown by blue trace), combined with corresponding multi-peak fit (dashed trace), substantiates previous analysis that the best sensitivity occurs when $\Omega_{\textrm{L}}\sim \Gamma_{\textrm{EIT}}$, with $\Delta_c=0$ corresponding to the half maximum of EIT lines associated with each peak. Fixing $E_\textrm{L}=3.0$ mVcm$^{-1}$ and $\delta_s=150.000$ kHz, we use quantum superhet to measure MW electric field below.

We first demonstrate the slope detection of quantum superhet. By measuring $P_s$ for a measurement time of $1$ s for various $E_s$, we present the data in Fig.~\ref{fig:3} (c) (light blue trace with filled circles). We see that $P_s\propto E_s$ for $E_s\ll E_\textrm{L}$, with nonlinear behavior only becoming visible for $E_s\sim E_{\textrm{L}}$, as expected. (The linear dynamical range is $90$ dB). The sensitivity $\mathcal{S}$ can be readily obtained as $\mathcal{S}=55$ nVcm$^{-1}$Hz$^{-1/2}$ ($-145$ dBVcm$^{-1}$Hz$^{-1/2}$), with signals below sinking into the noise base (purple block). 

Figure~\ref{fig:3} (c) moreover shows that - due to the slope detection - quantum superhet has superior sensitivity in comparison with Rydberg-atom MW electrometers performing nonlinear detection (by two orders of magnitude). Via small modifications of our setup~\cite{footnote2}, we have realized a room-temperature Rydberg-atom electrometer~\cite{Sedlacek2012,Fan2015}. To avoid low-frequency noises, we have added a $100\%$ amplitude modulation at $150.000$ kHz to the MW signal, and demodulated with a spectrum analyzer with $1$ Hz RBW. We then measure changes $\Delta P(E_s)=P(E_s)-P(0)$ of resonant EIT transmission with and without signals (dark blue trace with filled squares). Linear fitting of the experimental data shows $\Delta P(E_s) \propto E_{s}^{1.8}$ (i.e., nearly quadratic) for $\Omega_s\ll \Gamma_{\textrm{EIT}}\approx\Omega_{\textrm{L}}$, with obvious disadvantage that the unit signal-to-noise ratio (SNR) is reached at significantly larger $E_s$ than quantum superhet. Using this electrometer, we achieve a sensitivity of $30$ $\mu$Vcm$^{-1}$Hz$^{-1/2}$ ($-90$ dBVcm$^{-1}$Hz$^{-1/2}$). While state of the art sensitivity of such electrometer reaches $3$ $\mu$Vcm$^{-1}$Hz$^{-1/2}$~\cite{Kumar2017} in the photon shot noise limit, achieving a sensitivity comparable to quantum superhet is very hard, as highly squeezed lights are required to overcome the optical limit. Note for $\Omega_s>\Gamma_{\textrm{EIT}}$, the AT-splitting measurement (gray trace) also detects $E_s$ linearly, but $\Gamma_{\textrm{EIT}}$ sets a constraint on the sensitivity. 

According to Fig.~\ref{fig:3} (d), we estimate the smallest detectable MW field as $E_{\textrm{min}}=2.4$ nVcm$^{-1}$ for a measurement time $T=2097.1$ s corresponding to $0.5$ mHz RBW of the spectrum analyzer. We have achieved this by reducing several technical noises affecting the long time stability of our system (Supplementary material), which leads to SNR of $27.5$ dB for $E_s=55$ nVcm$^{-1}$. Since slope detection renders the scaling $E_{\textrm{min}}=\mathcal{S}/\sqrt{T}$, we theoretically estimate $E_{\textrm{min}}=1.2$ nVcm$^{-1}$ for $T=2097.1$ s. The deviation of experimental result from the theory may be attributed to the fact that our spectrum analyzer is not synchronized with GPSDO, which leads to a frequency drift between the two and thus the $6$ dB loss. Since $T$ is in principle limited by the frequency stability of GPSDO and can be extended to more than $4$ hours using recently developed techniques~\cite{Boss2017}, we can achieve $E_{\textrm{min}}\lesssim 0.5$ nVcm$^{-1}$. 

The remarkably high sensitivity ensures the phase and frequency of a MW signal field to be detected at high accuracies, as shown below. We first experimentally demonstrate quantum sensing of a MW signal in the form $\Omega_{\textrm{in}}(t)=\sum_{i=1,2}(r_ie^{-i\delta^i_dt})\Omega_te^{-i\delta_st}$, with the goal to resolve frequency $\delta_d^i$. This is motivated by the active radar detection~\cite{Doviak2012}, where $\Omega_te^{-i\delta_st}$ is reflected from different moving objects with reflection rate $r_i<1$ and Doppler frequency shift $\delta^i_d$, which information allows identification of velocities. Creating the signal with two antennas and using the spectrum analyzer with a RBW of $1$ Hz, we analyze $P_\textrm{out}(t)$ in the frequency domain [Fig.~\ref{fig:4}(a)]. First taking $\delta_d^1=0$, $\delta_d^2=-1$ Hz, and $r_i\Omega_t=20$ kHz (corresponding a field strength of $7.8$ $\mu$Vcm$^{-1}$), we observe two equal-hight peaks at frequencies $150.000$ kHz and $149.999$ kHz (red trace), whose relative shift from the reference case $\delta^i_d=0$ provides measurement of $\delta^i_d$. Provided the inter-peak dip is above $3$ dB, the frequency resolution is limited by RBW of the spectrum analyzer, i.e., $1$ Hz here. To exam the frequency accuracy, we choose $\delta_d^1=0$ and $\delta_d^2=4$ Hz when two peaks are well separated (blue trace), where the frequency accuracy can be obtained by $\textrm{RBW}/\sqrt{2\times\textrm{SNR}}$. It follows from Fig.~\ref{fig:3} (d) that a frequency accuracy of $30$ $\mu$Hz can be achieved for such a small signal of $55$ nVcm$^{-1}$ for $T=2097.1$ s~\cite{footnote3}. 

We then demonstrate experimental detection of phase $\phi_s$ in a MW signal $\Omega_se^{-i(\delta_st+\phi_s)}$. For $\Omega_s=20$ kHz, we extract a phase $\phi_{\textrm{out}}$ from $P_{\textrm{out}}(t)$, using a lock-in amplifier synchronized by GPSDO with $1.04$ Hz equivalent noise bandwidth (Supplementary material). Figure~\ref{fig:4} (b) compares $\phi_{\textrm{out}}$ and $\phi_s$,  showing good agreement between the two. The inset shows experimental data of $\phi_{\textrm{out}}$ (black trace) for an input phase $\phi_{\textrm{s}}$ which jumps by $1$ degree in a stepwise fashion (red trace). By measuring the standard deviation of the phase fluctuation, we estimate the phase resolution in our experiment as $0.8$ degree, agreeing with the theory; see supplementary material.

Concluding, we have developed a novel technique for phase- and frequency-resolved quantum sensing of MW electric fields in experiments. From a fundamental standpoint, realization of favorable scalings of sensitivity for ultraweak fields is remarkable: It pushes forward the limit of atom-based MW electric field quantum sensing without using non-classical resources, with a clear roadmap to improve present modest setup towards quantum projection noise limited (QPNL) sensitivity $\sim 700$ pVcm$^{-1}$Hz$^{-1/2}$ (Supplementary material), ensuring ultrahigh measurement accuracies of frequency and phase. Equally appealing are outstanding experimental simplicity of quantum superhet, its general applicability to sense electric field from radio frequency to the far infrared, and feasibility for miniaturization and integrability~\cite{Bajcsy2009,Kuebler2010,Ghosh2006}. Our work provides a remarkable step en route to realizing future quantum receivers, such as in radars or radio telescopes with the benefit of SI-traceable accuracy and ultrahigh sensitivity~\cite{Holloway2014,Holloway2017}, or in terahertz communication~\cite{Koenig2013,Nagatsuma2016}, allowing to recover the information encoded in terahertz carriers via phase or frequency modulations. 

The authors would like to acknowledge discussions with Benoit Vermersch and Hannes Pichler. This research is funded by National Key R\&D Program of China (Grant No. $2017$YFA$0304203$), National Natural Science Foundation of China (Grant No. $61827824$, $61527824$, $11874038$), $1331$ Key Subjects Construction, and 111 project (Grant No. D$18001$). L.T.X acknowledges support from the 
Program for Changjiang Scholars and Innovative Research Team (Grant No. IRT$13076$). Y.H. acknowledges support from the National Thousand-Young-Talents Program.

\makeatletter


\newpage

\section{Supplemental material}

\subsection*{Theory}

We derive the EIT transmission for the quantum superhet based on the setup in Fig.~\ref{fig:1}(a) under resonant conditions $\Delta_{p/c}=0$ for both coupling and probe lasers. The relevant Hamiltonian takes the form (in the basis of bare states $[|1\rangle, |2\rangle, |3\rangle, |4\rangle]^T$)
\begin{equation}\label{eq:H}
H(t)=\hbar\left(
\begin{array}{cccc}
0& \frac{\Omega_p}{2} & 0 & 0 \\
 \frac{\Omega_p}{2} &0 & \frac{\Omega_c}{2} & 0 \\
0 & \frac{\Omega_c}{2} &0& \frac{\Omega_{\textrm{L}}+e^{-iS(t)}\Omega_s}{2}\\
0 & 0 & \frac{\Omega_{\textrm{L}}+e^{iS(t)}\Omega_s}{2} & 0 \\
\end{array}
\right).
\end{equation}
Here $S(t)$ denotes the time-dependent relative phase $S(t)=\delta_s t+\phi_s$ between the signal and local MW fields. 

Accounting for the spontaneous emission, the dynamics of our system is described by the master equation for density matrix $\dot{\rho}$, i.e.,
\begin{equation}
\dot{\rho}=\frac{i}{\hbar} [\rho,H(t)]+\mathcal{D}[\rho], \label{eq:Master}
\end{equation}
where the second term is explicitly written as 
\begin{equation}
\mathcal{D}[\rho]\equiv\left(
\begin{array}{cccc}
\gamma_2\rho_{22}+\gamma_4\rho_{44}& -\frac{\gamma_2}{2}\rho_{12} & -\frac{\gamma_3}{2}\rho_{13} &  -\frac{\gamma_4}{2}\rho_{14} \\
 -\frac{\gamma_2}{2}\rho_{21}  & \gamma_3\rho_{33}-\gamma_2\rho_{22} & -\frac{\gamma_{23}}{2}\rho_{23}  & -\frac{\gamma_{24}}{2}\rho_{24} \\
 -\frac{\gamma_3}{2}\rho_{31} &  -\frac{\gamma_{23}}{2}\rho_{32} & - \gamma_3\rho_{33} & -\frac{\gamma_{34}}{2}\rho_{34}\\
   -\frac{\gamma_4}{2}\rho_{41} & -\frac{\gamma_{24}}{2}\rho_{42} & -\frac{\gamma_{34}}{2} \rho_{43}& -\gamma_4\rho_{44}
\end{array}
\right).\label{eq:L}
\end{equation}
Here $\gamma_{\textrm{ij}}=(\gamma_{\textrm{i}}+\gamma_{\textrm{j}})$, where $\gamma_{i}$ ($i=2,3,4$) is the decay rate [Fig.~\ref{fig:1}(a)]. In writing Eq.~(\ref{eq:L}), we have ignored the spontaneous emission associated with $|3
\rangle-|4\rangle$ and other possible transitions, as they are comparatively small.  We are interested in the limit where $\delta_s$ in Eq.~(\ref{eq:H}) is small compared to all characteristic energy scales of the system dynamics. 

We first illustrate the key physics taking the example of cold atoms. Within the adiabatic approximation, the probe laser transmission associated with the instantaneous steady state is written in terms of the imaginary component of susceptibility as~\cite{Fleischhauer2005}
\begin{equation}
P(t)=P_ie^{-kL \Im[\chi(t)]}. \label{eq:P}
\end{equation}
Here $P_i$ is the incident light power,  $L$ is the length of the cell containing Rydberg atoms, $k=2\pi/\lambda_{\textrm{P}}$ is the wave-vector of probe laser. Note $\chi(t)=C\rho_{21}$ is the susceptibility associated with the instantaneous steady state, where $\rho_{21}(t)$ denotes the instantaneous steady-state density matrix component associated with $|1\rangle-|2\rangle$ transition. Furthermore $C=-2 N_0\mu_{12}^2/(\epsilon_0 \hbar \Omega_p)$, where $N_0$ is the total density of atoms, $\mu_{12}$ is the dipole moment of the ground state transition, and $\epsilon_0$ is the vacuum permittivity. 

For $\Omega_s\ll \Omega_{\textrm{L}}$, an analytical expression for $P(t)$ can be derived as follows [c.f. Eq.~(\ref{eq:EIT}) in the main text]. Assuming the ideal case where $\gamma_{3(4)}=0$, the imaginary part of the susceptibility $\chi(t)$ can be straightforwardly derived as
\begin{equation}
 \Im[\chi(\Omega, t)]=\chi_0 \frac{|\Omega|^2}{|\Omega|^2+\Gamma^2 }, \label{eq:rho12}
\end{equation}
where $\Gamma=\Omega_p\sqrt{\frac{2(\Omega_c^2+\Omega_p^2)}{(2\Omega_p^2+\gamma_2^2)}}$ is intimately related to the EIT linewidth, $\chi_0=\frac{C\gamma_2 \Omega_p}{\gamma_2^2+2\Omega_p^2}$ is the peak value of the spectrum, and $\Omega=|\Omega_{\textrm{L}}+e^{-iS(t)}\Omega_s|$. Perturbative expansions of Eq.~(\ref{eq:rho12}) in terms of the small parameter $\Omega_s/\Omega_{\textrm{L}}$ reads at the first order as 
\begin{eqnarray}
 \Im[\chi(\Omega_\textrm{L},t)]=\Im[\chi(\Omega_\textrm{L})]+S_\textrm{L}\Omega_s\cos(\delta_st+\phi_s). \label{eq:rho21}
\end{eqnarray}
Here $S_\textrm{L}=2\chi_0\left[\frac{\Gamma^2\Omega_{\rm{L}}}{(\Omega_{\rm{L}}^2+\Gamma^2)^2}\right]$ is the slope of spectrum (\ref{eq:rho12}) at $\Omega=\Omega_{\textrm{L}}$. When $\Omega_\textrm{L}=\Gamma/\sqrt{3}$, the spectrum is linear near $\Omega=\Omega_{\textrm{L}}$, corresponding to the maximum slope $S_{\textrm{max}}=3\sqrt{3}\chi_0/(8\Gamma)$. Let us denote $\alpha=kL\Gamma S_{\textrm{max}}$. Substituting Eq.~(\ref{eq:rho21}) into Eq.~(\ref{eq:P}), for $\alpha\Omega_s/\Gamma\ll 1$, we arrive at Eq.~(\ref{eq:EIT}) in the main text, with $\bar{P}=P_i e^{-kL\chi_0/4}$. 

The form of Eq.~(\ref{eq:rho21}) holds as well for thermal atoms, where the Doppler average of $\rho_{21}(t)$ is required for calculating $\Im[\chi(t)]$ in Eq.~(\ref{eq:P}), and when considering $\gamma_{3(4)}\neq 0$. Note these effects will lead to modified $\Gamma$ and $\alpha$. This way, we also obtain the theoretical value of the probe laser transmission shown in Figs.~(\ref{fig:1}) (a) and (b).

\subsection*{Experimental setup}

In our experiment, we use Cs atoms in a vapor cell at room-temperature. The cell is $5$-cm-long and contains ground-state atoms with a total density $N_0=4.89\times10^{10}$~cm$^{-3}$. We realize the four-level configuration in Fig.~\ref{fig:1} using four states in a Cesium atom: $6\textrm{S}_{1/2}$, $\textrm{F}=4$; $6\textrm{P}_{3/2}$, $\textrm{F}=5$; $47\textrm{D}_{5/2}$, and $48\textrm{P}_{3/2}$. The hyperfine states $6\textrm{S}_{1/2}$, $\textrm{F}=4$ and $6\textrm{P}_{3/2}$, $\textrm{F}=5$ comprise the lowest two states $|1\rangle$ and $|2\rangle$ in the configuration, with $\gamma_2=5.2$ MHz. Moreover, the Rydberg state $47\textrm{D}_{5/2}$, with inverse lifetime $\gamma_3=3.9$ kHz , and Rydberg state $48\textrm{P}_{3/2}$, with inverse lifetime $\gamma_4=1.7$ kHz, make up the states $|3\rangle$ and $|4\rangle$ there. In calculating $\gamma_3$ and $\gamma_4$ at room temperatures, we have considered black-body induced transitions up to $n=70$. We apply a local MW field at $6.94$ GHz to resonantly drive the Rydberg transition $47\textrm{D}_{5/2}\rightarrow 48\textrm{P}_{3/2}$. In detecting a MW signal, the local and the signal fields are combined by a 2-way microwave resistive power divider, and are coupled to free space via the same resonant horn antenna. The resonant coupling between hyperfine states $6\textrm{S}_{1/2}$, $\textrm{F}=4\rightarrow6\textrm{P}_{3/2}$, $\textrm{F}=5$ is realized using a $852$ nm probe beam provided by a commercial extended cavity diode laser (ECDL). The resonant coupling between states $6\textrm{P}_{3/2}$, $\textrm{F}=5$ and $47\textrm{D}_{5/2}$ is realized using a $510$ nm beam generated by a frequency-double diode laser. The probe and coupling laser beams counter propagate through the room-temperature Cs cell, with minimized Doppler broadening of the transition. Their polarizations are linear, and are parallel to the direction of MW fields, leading to excitations of the magnetic sub-level $|m|=1/2$. For the probe beam, the $1/e^2$ beam diameter is $1.70 \pm 0.04$ mm, and the optical power incident to the vapor cell is $120 \pm 4~\mu$W, yielding effectively ${\Omega_p=5.7 \pm 0.6}$ MHz. For the coupling beam, the $1/e^2$ beam diameter is $2.00 \pm 0.05$ mm, and the incident optical power is $34 \pm 1$ mW, yielding $\Omega_c=0.97\pm 0.12$ MHz. After absorption by Cs atoms, the power of the probe light incident on the detector is about $10$ $\mu$W.

\subsection*{Reduction of technical noise}

In our experiment, the $150.000$ kHz signal is analyzed by a spectrum analyzer. The frequency noise of the probe laser and the seed of coupling laser are actively canceled by locking them to a $10$-cm-long ultra-low expansion (ULE) glass cavity with frequency noise server (FNS). The cavity is double coated at $852$ nm and $1020$ nm with a finesse of $200000$. The linewidth of the high finesse cavity is about $7.5$ kHz. The cavity is placed in a vacuum system at a residual pressure below $10^{-8}$ mbar, and its temperature is stabilized to the zero crossing point of the coefficient of thermal expansion. The system is mounted on a passive vibration isolation platform, and is surrounded by the acoustic and temperature insulation box. The FNSs are realized by PDH technique, and the feedbacks are injected to the PZTs and the diode currents of both lasers. The locking bandwidths of the $852$ nm laser and the $1020$ nm laser are about $250$ kHz and $350$ kHz, respectively. The beat note result of the $852$ nm laser with another equal system shows that the linewidth of the 852 nm laser is below $20$ Hz. The linewidth of the $510$ nm laser is estimated to be $<40$ Hz. Low frequency intensity noises of both lasers are actively eliminated through a feedback to double-pass acousto-optic modulators (AOMs) in cateye configuration. The $852$ nm probe light is separated into two orthogonally polarized output beams using calcite beam displacer, which propagate in parallel through the center of the Cs vapor cell. High frequency common mode intensity noise of the probe light is canceled by means of the balanced detection technique. Both MW sources are synchronized with a GPS disciplined oscillator with Rubidium timebase (GPSDO), so as to minimize their long time frequency drift. The signal MW source is $150$.000 kHz detuned from the local MW source, which offsets the interference signal of the quantum superhet to a sufficient high frequency, preventing the low frequency electronic pink noise. 

\subsection*{Measuring the phase from EIT signal}

We extract a phase $\phi_{\textrm{out}}$ from $P_{\textrm{out}}(t)$ by using a lock-in amplifier. The filter slope of the lock-in amplifier is set to $18$ dB/oct and the time constant is fixed at $100$ ms to realize $1.04$ Hz equivalent noise bandwidth. This leads to a SNR of $44$ dB for $E_s=7.8$ $\mu$Vcm$^{-1}$, leading to theoretical estimation of the phase resolution as $0.6$ degree. To assess the phase resolution experimentally, the standard deviation of the fluctuation in $\phi_{\textrm{out}}$ is measured in a period of $1-5$ s and $6-10$ s, respectively, with $1$ s waiting time for the signal to reach 99\% of its final value.

\renewcommand{\bibnumfmt}[1]{[S#1]}
\renewcommand{\citenumfont}[1]{S#1}

 \begin{figure}[tb]
\centering
\includegraphics[width= 0.98\columnwidth]{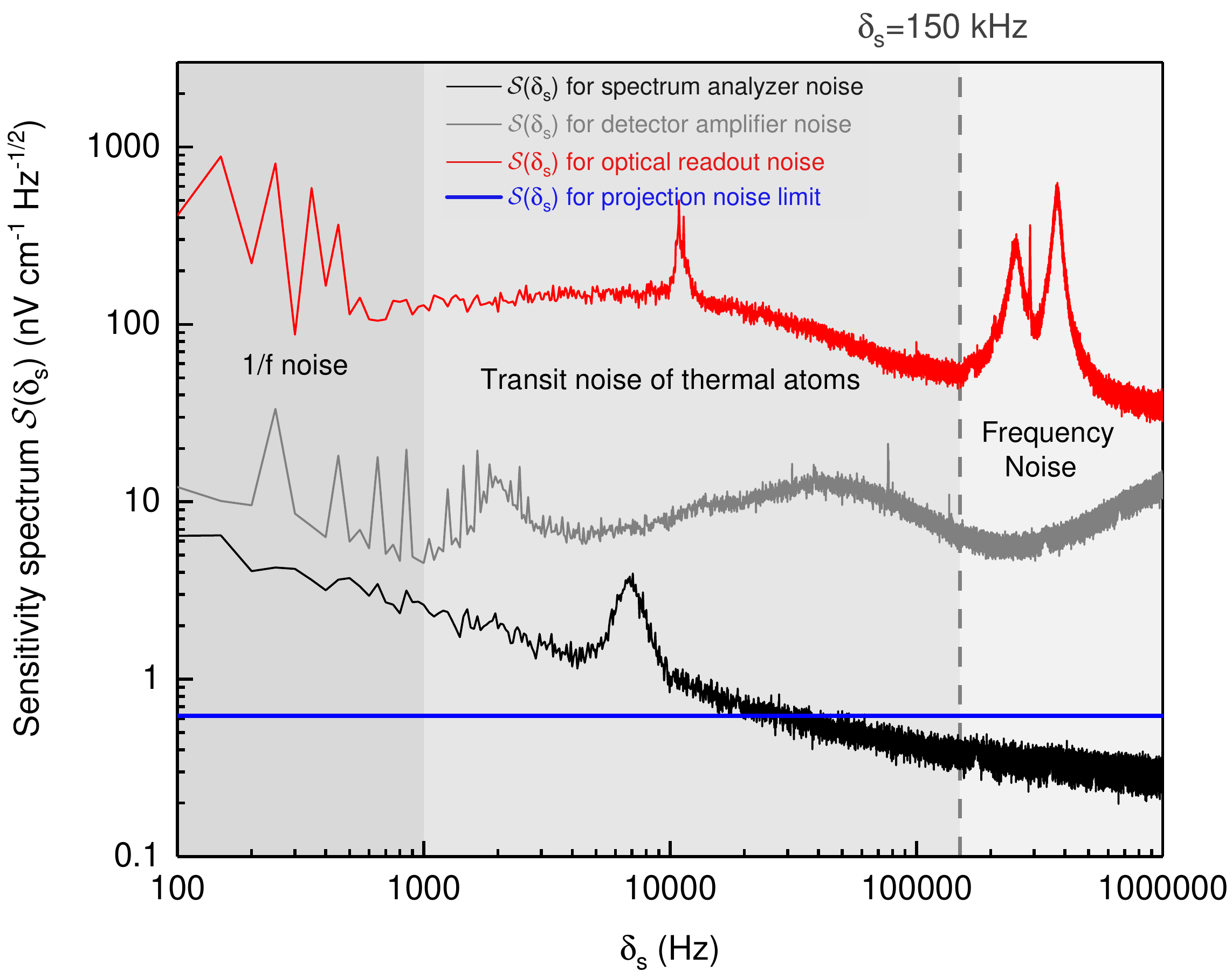}
\caption{Sensitivity $\mathcal{S}$ of quantum superhet as a function of $\delta_s$ of the signal MW electric field. Shown are respectively the sensitivity spectrum $\mathcal{S}(\delta_s)$ that corresponds to various optical readout noises (red), the amplifier noise of photon detector (gray) and the spectrum analyzer noise (black), and blue curve represents the QPNL sensitivity of our setup; see details in supplementary material. Note $\delta_s=150$ kHz is chosen in our experiments.}\label{fig:5}
\end{figure} 

\subsection*{Sensitivity spectrum}

In this section, we show how we obtain the sensitivity spectrum presented in Fig.~\ref{fig:5}. Let us denote by $\mathcal{S}(\delta_s)$ the sensitivity spectrum, i.e., the field sensitivity $\mathcal{S}$ at frequency $\delta_s$. Further, we let $S_{\textrm{P}}(\delta_s)$ denote the noise spectrum associated with $P_{\textrm{out}}(t)$. Importantly, according to the linear relation in Eq. (2) of the main text, we can write $\mathcal{S}(\delta_s)=\kappa S_{\textrm{P}}(\delta_s)$, with $\kappa$ being a constant coefficient. 

To obtain $S_{\textrm{P}}(\delta_s)$, we use the relation $S_\textrm{P}(\delta_s)=[S_\textrm{PD}(\delta_s)\times R]^{1/2}/(G\mu)$. Here $S_{\textrm{PD}}(\delta_s)$ is the noise power density associated with each noise sources in our detection system including optical readout noises, the amplifier noise of photon detector, and the noise of the spectrum analyzer, $G$ and $\mu$ denote the trans-impedance gain and the response of detectors, respectively, and $R$ labels the impedance of the spectrum analyzer. In our setup, we have $G=175\times10^3$ V/A, $\mu=0.58$ A/W, $R=50$ $\Omega$. Morever, we can experimentally measure the noise power density $S_{\textrm{PD}}(\delta_s)$ for each aforementioned noise, thus obtains corresponding $S_{\textrm{P}}(\delta_s)$.

To determine the coefficient $\kappa$, we note that the sensitivity of $55$ nVcm$^{-1}$Hz$^{-1/2}$ is achieved at $\delta_s=150.000$ kHz, where we have measured $S_{\textrm{PD}}=-110$ dBm/Hz associated with the optical readout noise. This gives $\kappa=7.9\times10^3$ Vcm$^{-1}$W$^{-1}$. 

Combinations $S_{\textrm{PD}}(\delta_s)$ for each type of noises above and knowledge of $\kappa$, we plot all the sensitivity spectra shown in Fig.~\ref{fig:5}. 

 \subsection*{Roadmap to QPNL sensitivity}
 
We first present a detailed noise analysis based on the sensitivity spectrum in Fig.~\ref{fig:5}. We see that the primary source of noise limiting the sensitivity of quantum superhet varies with $\delta_s$ of the signal: For frequencies below $1$ kHz, the 1/f noise of the electric circuits dominates over other noises; Between $1$ kHz and $100$ kHz, the transit noise due to thermal atoms provides the main noise source; For frequencies above $100$ kHz, it is the frequency noise of coupling and probe lasers caused by FNS resonant that mainly limits the sensitivity. In view of the requirement of both optimal sensitivity and $\delta_s\ll \Gamma_{\textrm{EIT}}$, we choose $\delta_s = 150.000$ kHz for our experimental demonstration. 

Now we determine the QPNL sensitivity for our setup. Since the quantum superhet is operated in slope detection mode, the QPNL sensitivity is formally given by 
\begin{equation}
E_{\textrm{QPNL}}= \frac{\sqrt{2}\hbar}{2\mu_{r}}\frac{1}{\sqrt{N_{a}}\tau_c}.
\end{equation}
Here $\mu_{r}$ is the dipole moment associated with Rydberg transition, $\tau_c=1/\Gamma$ is the coherence time of the quantum superhet, and $N_{a}$ is the atom number participating in EIT process per second. For our setup, we estimate $\tau_c$ from $\tau_c\approx \Gamma^{-1}_{\textrm{EIT}}$, with $\Gamma_{\textrm{EIT}}=7.9$ MHz from experiment results. The $N_{a}$ is estimated as $2.14\times 10^{13}$ s$^{-1}$ for EIT process, which leads to an enhancement of $5$ $\mu$W light transmission at the photon detector compared to the case without coupling laser. This gives $E_{\textrm{QPNL}}=700$ pV$\textrm{cm}^{-1}$ $\textrm{Hz}^{-1/2}$ indicated in Fig.~\ref{fig:5}.

Finally we outline the roadmap toward the QPNL sensitivity according to Fig.~\ref{fig:5}. For $\delta_s= 150.000$ kHz, we see that the primary noises sources in our detection scheme can be systematically eliminated using techniques feasible within present quantum sensing experiments as follows: (i) The transit noise due to thermal atoms can be eliminated by using larger-diameter probe and coupling beams; (ii) The laser frequency noise can be readily eliminated by using state of the art lasers with mHz linewidth~\cite{Kessler2012,Matei2017} and by expanding servo bandwidth to several MHz~\cite{Musha1997,Endo2018}; (iii) The amplifier noise of photon detector can be reduced by means of optical heterodyne or homodyne detection~\cite{Cox2018,Kumar2017a}; (iv) The spectrum analyzer noise can be removed by using conventional electronic amplifiers. After the relevant technical noises have been eliminated, quantum superhet approaches QPNL. \\

\subsection*{Calibration of MW electric field amplitude}

In our experiment, the signal and local MW E-fields are emitted from difference sources, respectively. To calibrate each field, we follow the procedures below. We apply a test MW field to resonantly drive the Rydberg transition in the 4-level EIT configuration. We denote the test field amplitude by $E$, chosen to be sufficiently large to ensure its subsequent measurement to a high accuracy. First, from the output power of MW source, one can calculate $E$ according to the standard antenna equation (IEEE Std 1309-2013), i.e., 
\begin{equation}
E=\sqrt{\frac{\eta(P_s-\alpha_l)g}{4\pi d^2}}.\label{eq:IEEE}
\end{equation}
Here, $\eta=377$ $\Omega$ is the intrinsic impedance of free space, $P_s$ is the output power of MW source, $\alpha_l$ is the insertion loss between MW source and antenna, $g$ is the gain of antenna, and $d$ is the distance between the transmitting antenna and the receiving point. Determination of $\alpha_l$ requires the experimental data of $E$, measured via the AT-splitting approach, where we read off $E$ from the relation $E=\sqrt{2}\pi\Delta_{\textrm{AT}}\hbar/\mu_{r}$. In the end, the experimental data for $E$ combined with Eq.~(\ref{eq:IEEE}) allow us to calibrate the insertion loss $\alpha_l$ for both the local and signal MW fields. For the local MW field, we find $\alpha_l=14.7$ dB. This includes the $6.5$ dB RPD insertion loss and four meters $1.5$ dB/m wire loss, while the remaining insertion loss can be attributed to the connectors insertion loss. For the signal field, we have obtained $\alpha_l=11.9$ dB, which is smaller than the local MW field due to utility of a shorter transmission wire ($3$ m). Once the insertion loss has been calibrated, the field strengths for both the signal and local MW fields can be readily determined using Eq.~(\ref{eq:IEEE}).

\clearpage 


\medskip

\medskip

\begin{thebibliography}{1}
\bibitem{Spitler2016}
L.G. Spitler, P. Scholz, J.W. T. Hessels, S. Bogdanov, A. Brazier, F. Camilo, S. Chatterjee, J.M. Cordes, F. Crawford, J. Deneva, et~al. A repeating fast radio burst. \textit{Nature} \textbf{531}, 202–-205 (2016).
\bibitem{Shannon2018}
R.M. Shannon, J.P. Macquart, K.W. Bannister, R.D. Ekers, C.W. James, S. Os{\l}owski, H. Qiu, M. Sammons, A. W. Hotan, M.A. Voronkov, et~al. The dispersion--brightness relation for fast radio bursts from a  wide-field survey. \textit{Nature} \textbf{562}, 386--390 (2018).
\bibitem{Alsdorf2000}
D.E. Alsdorf, J.M. Melack, T. Dunne, L.A. K. Mertes, L.L. Hess, and L.C. Smith. Interferometric radar measurements of water level changes on the Amazon flood plain. \textit{Nature} \textbf{404}, 174--177  (2000).
\bibitem{Rignot2008}
E. Rignot, J.L. Bamber, M.R. Van Den Broeke, C. Davis, Y. Li, W.J. Van De Berg, and E.V. Meijgaard. Recent Antarctic ice mass loss from radar interferometry and regional climate modelling. \textit{Nature Geoscience} \textbf{1}, 106--110 (2008).
\bibitem{Rohde2017}
U.L. Rohde and J.C. Whitaker. \textit{Communications Receivers: Principles and Design, Fourth  Edition}. McGraw-Hill Education, 2017.
\bibitem{Maeda2005}
H. Maeda, D.V.L. Norum, and T.F. Gallagher. Microwave manipulation of an atomic electron in a classical orbit. \textit{Science} \textbf{307}, 1757--1760 (2005).
\bibitem{Bienfait2017}
A. Bienfait, P. Campagne-Ibarcq, A. H. Kiilerich, X. Zhou, S. Probst, J. J. Pla, T. Schenkel, D. Vion, D. Esteve, J. J.  L. Morton, et~al. Magnetic resonance with squeezed microwaves. \textit{Phys. Rev. X} \textbf{7}, 041011 (2017).
\bibitem{Fan2015}
H. Fan, S. Kumar, J.A. Sedlacek, H. K\"{u}bler, S. Karimkashi, and J.P. Shaffer. Atom based RF electric field sensing. \textit{Journal of Physics B: Atomic, Molecular and Optical Physics} \textbf{48}, 202001 (2015).
\bibitem{Sedlacek2012}
J.A. Sedlacek, A. Schwettmann, H. K\"{u}bler, R. L\"{o}w, T. Pfau, and J.P. Shaffer. Microwave electrometry with Rydberg atoms in a vapour cell using bright atomic resonances. \textit{Nature Physics} \textbf{8}, 819--824 (2012).
\bibitem{Kumar2017a}
S. Kumar, H. Fan, H. K\"{u}bler, J. Sheng, and J.P. Shaffer. Atom-based sensing of weak radio frequency electric fields using homodyne readout. \textit{Sci. Rep.} \textbf{7}, 42981 (2017).
\bibitem{Kumar2017}
S. Kumar, H. Fan, H. K\"{u}bler, A.J. Jahangiri, and J.P. Shaffer. Rydberg-atom based radio-frequency electrometry using frequency modulation spectroscopy in room temperature vapor cells. \textit{Optics Express} \textbf{25}, 8625--8637 (2017).
\bibitem{Facon2016}
A. Facon, E.K. Dietsche, D. Grosso, S. Haroche, J.M. Raimond, M. Brune, and S. Gleyzes. A sensitive electrometer based on a Rydberg atom in a Schrodinger-cat state. \textit{Nature} \textbf{535}, 262--265 (2016).
\bibitem{Koenig2013}
S. Koenig, D. Lopez. Diaz, J. Antes, F. Boes, R. Henneberger, A. Leuther, A. Tessmann, R. Schmogrow, D. Hillerkuss, R. Palmer, et~al. Wireless sub-THz communication system with high data rate. \textit{Nature Photonics} \textbf{7}, 977--981 (2013).
\bibitem{Nagatsuma2016}
T. Nagatsuma, G. Ducournau, and C.C. Renaud. Advances in terahertz communications accelerated by photonics. \textit{Nature Photonics} \textbf{10}, 371--379 (2016).
\bibitem{Degen2017}
C.L. Degen, F. Reinhard, and P. Cappellaro. Quantum sensing. \textit{Rev. Mod. Phys.} \textbf{89}, 035002 (2017).
\bibitem{Sedlacek2013}
J.A. Sedlacek, A. Schwettmann, H. K\"{u}bler, and J.P. Shaffer. Atom-based vector microwave electrometry using rubidium Rydberg atoms in a vapor cell. \textit{Phys. Rev. Lett.} \textbf{111}, 063001 (2013).
\bibitem{Holloway2014}
C.L. Holloway, J.A. Gordon, S. Jefferts, A. Schwarzkopf, D.A. Anderson, S.A. Miller, N. Thaicharoen, G. Raithel. Broadband Rydberg Atom-Based Electric-Field Probe for SI-Traceable, Self-Calibrated Measurements. \textit{IEEE Transactions on Antennas and Propagation} \textbf{62}, 6169-6182 (2014).
\bibitem{Holloway2017}
C.L. Holloway, M.T. Simons, J.A. Gordon, A. Dienstfrey, D.A. Anderson, and G. Raithel. Electric field metrology for SI traceability: Systematic measurement uncertainties in electromagnetically induced transparency in atomic vapor. \textit{Journal of Applied Physics} \textbf{121}, 233106 (2017).
\bibitem{Cox2018}
K.C. Cox, D.H. Meyer, F.K. Fatemi, and P.D. Kunz. Quantum-limited atomic receiver in the electrically small regime. \textit{Phys. Rev. Lett.} \textbf{121}, 110502 (2018).
\bibitem{Boehi2012}
P. B\"{o}hi and P. Treutlein. Simple microwave field imaging technique using hot atomic vapor cells. \textit{Appl. Phys. Lett.} \textbf{101}, 181107 (2012).
\bibitem{Wade2017}
C.G. Wade, N. {\v{S}}ibali{\'c}, N.R. de Melo, J.M. Kondo,  C.S. Adams, and K.J. Weatherill. Real-time near-field terahertz imaging with atomic optical fluorescence. \textit{Nature Photonics} \textbf{11}, 40-43 (2017).
\bibitem{Doviak2012}
C. Alabaster. \textit{Pulse Doppler Radar: Principles, Technology, Applications}. SciTech Publishing, 2012.
\bibitem{Pandey2018}
K. Pandey. Single reference atomic based MW interferometry using EIT.  arXiv:1802.09935 (2018).
\bibitem{Fleischhauer2005}
M. Fleischhauer, A. Imamoglu, and J.P. Marangos. Electromagnetically induced transparency: Optics in coherent media. \textit{Rev. Mod. Phys.} \textbf{77}, 633 (2005).
\bibitem{footnote2}{In our experiment, we remove the local MW field, and tune the weak MW signal to be resonant with Rydberg transitions.}
\bibitem{Boss2017}
J.M. Boss, K.S. Cujia, J. Zopes, and C.L. Degen. Quantum sensing with arbitrary frequency resolution. \textit{Science} \textbf{356}, 837--840 (2017).
\bibitem{footnote3}{Note that when applied to actual detection of the speed of moving objects, one can infer respective velocity from corresponding Doppler frequency shift by $\upsilon_i=c\delta_d^i/(2f_0)$. Here $c$ and $f_0$ denote the speed of light and carrier frequency, respectively. Thus our experimental results correspond to a velocity measurement accuracy of $0.7$ $\mu$m/s, with a velocity resolution of $20$ $\mu$m/s.}
\bibitem{Ghosh2006}
S. Ghosh, A.R. Bhagwat, C.K. Renshaw, S. Goh, A.L. Gaeta, and B.J. Kirby. Low-light-level optical interactions with rubidium vapor in a photonic band-gap fiber. \textit{Phys. Rev. Lett.} \textbf{97}, 023603 (2006).
\bibitem{Bajcsy2009}
M. Bajcsy, S. Hofferberth, V. Balic, T. Peyronel, M. Hafezi, A.S. Zibrov, V. Vuletic, and M.D. Lukin. Efficient all-optical switching using slow light within a hollow fiber. \textit{Phys. Rev. Lett.} \textbf{102}, 203902 (2009).
\bibitem{Kuebler2010}
H. K\"{u}bler, J.P. Shaffer, T. Baluktsian, R. L\"{o}w, and T. Pfau. Coherent excitation of rydberg atoms in micrometre-sized atomic vapour cells. \textit{Nature Photonics} \textbf{4}, 112--116 (2010).
 \end{thebibliography}

\medskip

\begin{thebibliography}{1}
\bibitem{Kessler2012}
T. Kessler, C. Hagemann, C. Grebing, T. Legero, U. Sterr, F. Riehle, M.J. Martin, L. Chen, and J. Ye. A sub-40-mHz-linewidth laser based on a silicon single-crystal optical cavity. \textit{Nature Photonics} \textbf{6}, 687–692 (2012).
\bibitem{Matei2017}
D.G. Matei, T. Legero, S.H{\"a}fner, C. Grebing, R. Weyrich, W. Zhang, L. Sonderhouse, J.M. Robinson, J. Ye, F. Riehle, et~al. 1.5 $\mu$m lasers with sub-10 mHz linewidth. \textit{Phys. Rev. Lett.} \textbf{118},263202 (2017).
\bibitem{Musha1997}
M. Musha, K. Nakagawa, and K. Ueda. Wideband and high frequency stabilization of an injection-locked Nd:YAG laser to a high-finesse Fabry--Perot cavity. \textit{Optics letters} \textbf{22}, 1177--1179 (1997).
\bibitem{Endo2018}
M. Endo and T.R. Schibli.
\newblock Residual phase noise suppression for Pound-Drever-Hall cavity stabilization with an electro-optic modulator. \textit{OSA Continuum} \textbf{1}, 116--123 (2018).
\bibitem{Cox2018}
K.C. Cox, D.H. Meyer, F.K. Fatemi, and P.D. Kunz. Quantum-limited atomic receiver in the electrically small regime. \textit{Phys. Rev. Lett.} \textbf{121}, 110502 (2018).
\bibitem{Kumar2017a}
S. Kumar, H. Fan, H. K\"{u}bler, J. Sheng, and J.P. Shaffer. Atom-based sensing of weak radio frequency electric fields using homodyne readout. \textit{Sci. Rep.} \textbf{7}, 42981 (2017).
 \end{thebibliography}
\end{document}